\documentclass{aa}                                         %%as stated before: choose the right one...%
\usepackage{color}                                                  %% FD: only for corrections!!!!!
\usepackage{array}                                                  %% FD call:a little bit more space in tables....
\usepackage{txfonts,epsfig,graphicx,natbib,url,twoopt}
\usepackage[breaklinks=true]{hyperref}                              %% to avoid \citeads line fills
\usepackage{multirow}
\usepackage{amsmath}
\bibpunct{(}{)}{;}{a}{}{,}    %% natbib cite format used by A&A and ApJ
\newcommandtwoopt{\citeads}[3][][]{\href{http://adsabs.harvard.edu/abs/#3}%
                                        {\citealp[#1][#2]{#3}}}
\newcommandtwoopt{\citepads}[3][][]{\href{http://adsabs.harvard.edu/abs/#3}%
                                         {\citep[#1][#2]{#3}}}
\newcommandtwoopt{\citetads}[3][][]{\href{http://adsabs.harvard.edu/abs/#3}%
                                         {\citet[#1][#2]{#3}}}
\newcommandtwoopt{\citeyearads}
  [3][][]{\href{http://adsabs.harvard.edu/abs/#3}%
  {\citeyear[#1][#2]{#3}}}
\begin{document}
\title{Confirming the thermal Comptonization model for black hole X-ray emission in the low-hard state}

   \subtitle{}

   \author{
          M. Castro
          \inst{\ref{inpe}}
          \and
          F. D'Amico
          \inst{\ref{inpe}}
          \and
          J. Braga
          \inst{\ref{inpe}}
          \and
          T. Maiolino
          \inst{\ref{ferrara}}
          \and
          K. Pottschmidt
          \inst{\ref{nasa}}
          \and
          J. Wilms
          \inst{\ref{remeis}}
          }
   \institute{
              Instituto Nacional de Pesquisas Espaciais (INPE), São José dos Campos, Brazil \\
              \email{[castro|damico|braga]@das.inpe.br} \label{inpe}
  	      \and
     	      Physics  Earth Sciences Department University of Ferrara, Ferrara (Italy) 
              {\&} Universit\'e de Nice Sophia-Antipolis, Nice (France)\\
     	      \email{maiolino@fe.infn.it}\label{ferrara} 
                \and
               Cresst, NASA/GSFC {\&} UMBC, USA       \\                               
              \email{katja@milkyway.gsfc.nasa.gov}\label{nasa}  
              \and
              Remeis Observatory {\&} ECAP, University of Erlangen-Nuremberg, Germany\\
              \email{joern.wilms@sternwarte.uni-erlangen.de}\label{remeis}
             }
\date{ }
\abstract{ 
          Hard X-ray spectra of black hole binaries in the low/hard
          state are well modeled by thermal Comptonization of soft seed
          photons by a corona-type region with 
          $kT$\thinspace$\sim 50${\thinspace}keV and optical depth around 1. 
          Previous spectral
          studies of 1E{\thinspace}1740.7$-$2942, including both the soft and
          the hard X-ray bands, were always limited by gaps in the spectra or
          by a combination of observations with imaging and non-imaging
          instruments.
          In this study, we have used three rare
          nearly-simultaneous observations of 1E{\thinspace}1740.7$-$1942
          by both XMM-Newton and INTEGRAL satellites to combine spectra
          from four different imaging instruments with no data gaps, and we
          successfully applied the Comptonization scenario to explain the
          broadband X-ray spectra of this source in the low/hard
          state. For two of the three
          observations, our analysis also shows that, models including 
	Compton reflection can adequately
          fit the data, in agreement with previous reports. We show that
          the observations can also be modeled by a more detailed
          Comptonization scheme. Furthermore, we find the presence of an iron K-edge absorption feature
	  in one occasion, which confirms what had been previously observed by Suzaku. Our broadband
          analysis of this limited sample shows a rich spectral
          variability in 1E{\thinspace}1740.7$-$2942 at the low/hard state,
          and we address the possible causes of these variations. More
          simultaneous soft/hard X-ray observations of this system and
          other black-hole binaries would be very helpful in constraining the
          Comptonization scenario and shedding more light on the physics of
          these systems.
          }
\keywords{
          Radiation mechanisms: non-thermal      --
          Radiation mechanisms: thermal          --
          X-rays              : stars
         }
\titlerunning{XMM and INTEGRAL observations of non-thermal emission of 1E{\thinspace}1740.7$-$2942}
\maketitle
%
%________________________________________________________________

\section{Introduction}
\label{intro}

Since its discovery with the Einstein Observatory
(\citeads{1984ApJ...278..137H}), the putative black-hole system
1E{\thinspace}1740.7$-$2942 has been extensively studied. Historical
observations (\citeads{1987Natur.330..544S, 1991A&A...252..172S}) have
established 1E{\thinspace}1740.7$-$2942 as the brightest hard X-ray
source ($E${\thinspace}$>${\thinspace}$20${\thinspace}keV) in the
direction of the Galactic Center. From the years of observation with
SIGMA/Granat, the source was discovered to have 3 spectral states
resembling those of Cyg{\thinspace}X-1
(\citeads{1991ApJ...383L..49S}).  Another heritage from the SIGMA
years was a detection of 1E{\thinspace}1740.7$-$2942 up to
$\sim${\thinspace}$500${\thinspace}keV
(\citeads{1993ApJ...407..752C}). The source was dubbed a {\it
  microquasar} after radio jets were observed
(\citeads{1992Natur.358..215M}). The intense optical extinction
toward the Galactic Center  still prevents the identification of
the counterpart of 1E{\thinspace}1740.7$-$2942 at optical and infrared
wavelengths, despite all efforts
(\citeads{DelSanto05,Smith2002,2000A&A...363..184M}). ASCA
(\citeads{1999ApJ...520..316S}) as well as Chandra
(\citeads{Gallo2002}) that were helpful in determining the column density
of hydrogen ($N_{\mathrm{H}}$). Recent INTEGRAL observations have
shown that the source can be clearly detected up to
$\sim${\thinspace}$600${\thinspace}keV (\citeads{Bouchet2009}). A
broadband study with Suzaku with a small spectral gap
($10$--$12${\thinspace}keV) has shown the presence for the first
time of an iron K-edge absorption in the 1E{\thinspace}1740.7$-$2942
spectra (\citeads{Reynolds2010}).

In general, the combined soft/hard X-ray spectra of
1E{\thinspace}1740.7$-$2942 are fitted by a combination of a thermal
component and a non-thermal powerlaw. These fits are useful in
constraining the spectral state of 1E{\thinspace}1740.7$-$2942 through
the powerlaw index (see, e.g.,
{\citeads{Remillard06}). Alternatively, a combination of two thermal
models can be used to explain the broadband spectrum of this source:
a soft component, which is associated with the accretion disk, and a hard
Comptonized component coming from a corona-type region which has
$kT$\thinspace}$\sim$50{\thinspace}keV and an optical depth around 1
(e.g., \citeads{Bouchet2009}). In this study, we show evidence that
modeling the spectra of 1E{\thinspace}1740.7$-$2942 in the latter way
provides a consistent picture. We also report a detection here (by
XMM) of an iron K-edge absorption feature, which confirms the Suzaku
results (\citeads{Reynolds2010}).

Recently, a NuSTAR + INTEGRAL study on 1E{\thinspace}1740.7$-$2942 was
published based on data of 2012 (\citeads{Natalucci2013}). The
spectrum, starting at $\sim$3\,keV and extending up to 250\,keV, was
fitted by a combination of a Comptonization model ({\tt compTT}) and
a soft component ({\tt diskbb}), which enables us to make useful
comparisons with our results. Finally, we highlight that this present
study is an extended and improved version of a previous work, where we
analyzed data for only two epochs (2003 and 2005 ) and only from
PN/XMM and IBIS/INTEGRAL (\citeads{Castro2012}).

%  \vspace{15mm}
%______________________________________________________________
\section{Data selection and analysis}
\label{data}

To test the thermal Comptonization model in the low/hard
state of 1E{\thinspace}1740.7$-$2942, we carried out a search in the
databases of XMM and INTEGRAL and looked for nearly simultaneous
observations. The best matches satisfying our criteria are presented
in Table{\thinspace}{\ref{tab1}}, which results in three observations. The
2003 observation was performed by the two satellites almost
simultaneously. Some hours of delay can be seen in the 2005 data set,
and in 2012 the data from the two satellites are not simultaneous
with days of delay in between the observations. 

\begin{table*}
\setlength{\extrarowheight}{1ex}
\caption{
         XMM and INTEGRAL simultaneous observations of 1E{\thinspace}1740.7-2942.
        }
\label{tab1}
\begin{center}
\begin{tabular}{cccccc}
\hline \hline
 &	        &	PN            &	MOS1	            &	JEMX	          & ISGRI	         \\ 
\hline
\multirow{2}{*}{2003}		
 & TStart(UTC)\tablefootmark{(a)} & 2003-09-11 23:05:36 & 2003-09-11 23:00:21 & 2003-09-11 22:49:29 
                                                                              & 2003-09-11 22:50:31      \\ 
 & Exposure(ks)	&	1.6	      &	8.0		    &	3.0		  &	9.0		 \\ 
\hline
\multirow{2}{*}{2005}	              &	TStart(UTC)	    & 2005-10-02 01:21:18 &2005-10-02 01:16:05		
 & 2005-10-02 11:07:19		      & 2005-10-02 02:48:43		                                 \\	
 & Exposure(ks)	&	16.0	      &	7.7		    &	2.2		  &	23.4		 \\ 
\hline
\multirow{2}{*}{2012}		      &	TStart(UTC)	    & 2012-04-03 08:32:12 & 2012-04-03 08:27:03	
 & 2012-04-07 09:01:18		      & 2012-04-07 05:59:52		                                 \\
 & Exposure(ks)	&	82.7	      &	34.1		    &	1.6		  &	7.7		 \\ 
\hline \hline
\end{tabular}
\end{center}
 \tablefoot{
            \tablefoottext{a}{Time expressed in yyyy-mm-dd hh:mm:ss as derived from each produced spectrum file.}
           }
\end{table*}

Data from XMM-Newton (\citeads{Jansen01})
cameras PN (\citeads{Struder2001}), and MOS1
(\citeads{Turner2001}) were reduced using standard procedures with SAS
(V. 12.0.1) (\url{http://xmm.esa.int/sas/}). INTEGRAL 
(\citeads{2003A&A...411L...1W}) data from IBIS
(\citeads{Ubertini2003}) and JEM-X (\citeads{2003A&A...411L.231L})
telescopes were treated using the recipes described in the
OSA{\thinspace}10.0 documentation
(\url{http://www.isdc.unige.ch/integral/}). We have also made use of
{\it XSPEC} (V. 12.8.0) in performing our spectral fits.  Data from PN
were constrained to the $\sim${\thinspace}$2$ to $12${\thinspace}keV
region, and MOS1 data were limited to the
$\sim${\thinspace}$2$--$10${\thinspace}keV band.  The lower energy
threshold is due to the low count rate and signal-to-noise ratio, S/N
 below $\sim$2\,keV for both cameras (PN and MOS1) according to
the pileup analysis. We also note that $\sim${\thinspace}$2${\thinspace}keV 
was the lower limit in energy used by ASCA 
(\citeads{1999ApJ...520..316S}) and Chandra on two occasions 
(\citeads{2001ApJ...548..394C}; \citeads{Gallo2002}). With our spectra 
also starting around 2{\thinspace}keV we left the hydrogen column 
density as a free parameter in our fits.

The energy range from $\sim${\thinspace}10 up to $\sim${\thinspace}$20${\thinspace}keV was
filled with the use of the JEM-X/INTEGRAL telescope, even though
the 1E{\thinspace}1740.7$-$2942 count rates are quite low in this band for this
instrument. The data presented here from $20${\thinspace} up to $200${\thinspace}keV
were collected by the IBIS telescope onboard INTEGRAL,
completing our broadband coverage. We have made use only of the 
ISGRI/IBIS data (\citeads{2003A&A...411L.141L}).

Fits to our spectra included the multiplicative {\tt const} and
{\tt phabs} components in {\it XSPEC} when  one accounts for the
difference in the counts for the four instruments and the other
for absorption by neutral material.  The normalization factors
were relative to the PN instrument, which has the highest  count
rate among the four instruments.

While the ASCA (\citeads{1999ApJ...520..316S}) and Chandra
(\citeads{Gallo2002}) studies on 1E{\thinspace}1740.7$-$2942 have made
use of an absorbed powerlaw to fit the $2$--$20${\thinspace}keV
spectra, the broadband ($2$--$200${\thinspace}keV) Suzaku study
(\citeads{Reynolds2010}) used a combination of models. On the other
hand, the INTEGRAL hard X-ray spectrum of 1E{\thinspace}1740.7$-$2942
has been fitted with a Comptonization model (\citeads{Bouchet2009}). 
 
We have  thus fitted our XMM and INTEGRAL broadband
($2$--$200${\thinspace}keV) spectra of 1E{\thinspace}1740.7$-$2942
with two components: a soft standard extended blackbody 
({\tt  diskbb}), which comes from the accretion disk and a Comptonized
component. We have made use of thermal Comptonization and other
variations. This component is probably coming from a larger region (a
{\it corona\/}). In this two-component model, we tied (as usual)
the temperature of the seed soft photons of the Comptonized component
to the disk average temperature (modeled as {\tt diskbb}).  To help us
in classifying the 1E{\thinspace}1740.7$-$2942 spectral state, we
have also made use of a classical exponential folded powerlaw to fit
our spectra (see details in \citeads{Remillard06}). We note that this
model had also been used before for 1E{\thinspace}1740.7$-$2942 (e.g.,
\citeads{Bouchet2009}).

%______________________________________________________________
\section{Results}
\label{results}

%Figure1Begin
%author     : Flavio D'Amico
%short title: Spectra of 1E 1740.7-2942
%source     : made to appear here in this study
%---
%from -> http://www.aanda.org/index.php?option=com_content&task=view&id=171&Itemid=173
%        For a two-column-wide plot, substitute figure by figure*.
%        don't forget to end it with end{\figure*} <- FD
%        there is a litte remind below (commented) explaining hot to produce side captions. <-FD
%---
%\begin{figure*}
%\sidecaption
%\resizebox{\hsize}{!}{\includegraphics[angle=270]{fig1.pdf}}
%---
\begin{figure*}[ht]
\includegraphics[angle=270,scale=1.1]{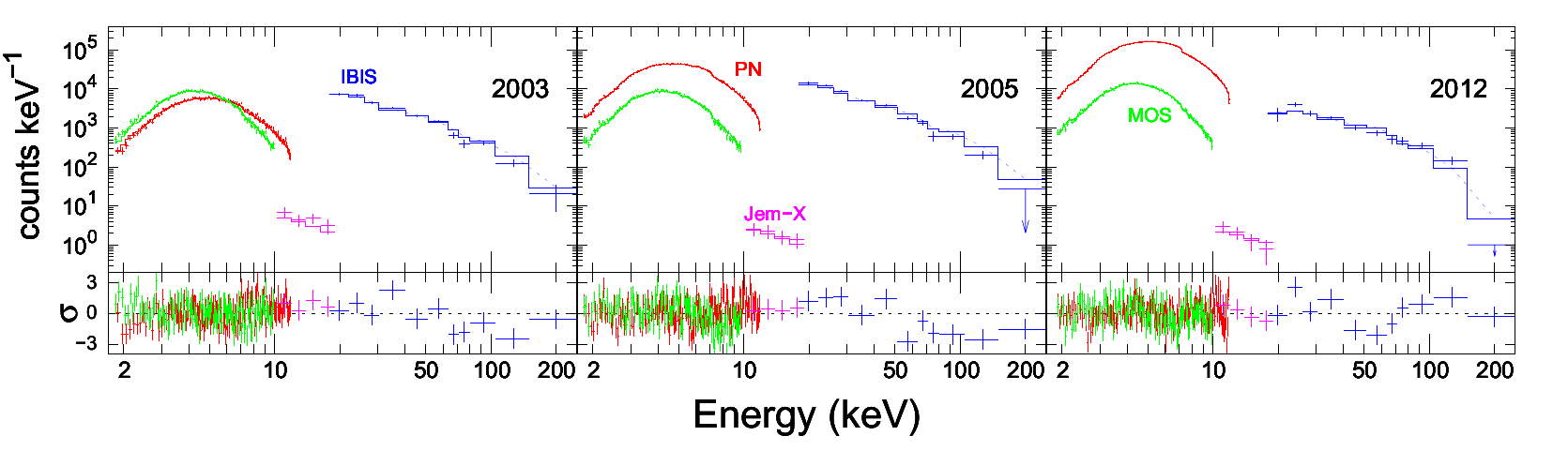}
         \caption{
                 XMM and INTEGRAL spectra of 1E{\thinspace}1740.7$-$2942 for 2003, 
                 2005, and 2012 observations described in this study for a fitted  
                 diskbb{\thinspace}$+${\thinspace}{\tt compTT} model. In the 2012
                 spectrum a Fe-edge in 7.1{\thinspace}keV (modeled as Gauss line: see text) 
                 is also present. Instruments are labeled in upper panels, which 
                 also shows the data, while bottom panels correspond to the residuals 
                 of the fits. In the middle and right upper panels, the bins at
                 $\sim${\thinspace}$200${\thinspace}keV are 3${\sigma}$ upper limits. 
                 In all the three upper panels, a dotted blue line shows the relative 
                 contribution of the {\tt compTT} component in the ISGRI part of the spectrum.
                 Fit parameters can be found in Table{\thinspace}{\ref{tab2}}.
                }
\label{fig1}
\end{figure*}
%Figure1End
%edbyFDend

The lower (XMM) parts of our spectra can be fitted very well as an
accretion disk consisting of multiple blackbody components, as the 
{\tt diskbb} model. It is noteworthy, however, that {\it another} (soft)
component is necessary (F$_{\hbox{\scriptsize{test}}}$ of the order of
10$^{-7}$) to adequately describe the soft part or our
1E{\thinspace}1740.7$-$2942 data in 2003 and 2012. This component is
only marginally needed in the 2005 spectrum
(F$_{\hbox{\scriptsize{test}}}$ of 10$^{-4}$).  Nevertheless, to
maintain uniformity in our comparisons between the three spectra, we kept
this second component in the XMM spectrum of 2005. This second
component is very well fitted by any Comptonization model. With the
addition of this Comptonization component, our broadband
$2$--$200${\thinspace}keV spectra can be very well modeled. As we have
already stated, instead of using Comptonization models, an exponential
folded powerlaw can also be used to fit (the INTEGRAL) part of our
spectra. We also performed those fits accordingly, which help us to
conclude (see details in \citeads{Remillard06})
that  1E{\thinspace}1740.7$-$2942 was in its (canonical) low/hard state
(LHS) in the three observations reported in this study.

It is interesting to note that we found no evidence
in any of our XMM data sets of the emission lines due to the
soft X-ray background in the direction of this source, as reported, for
example, in {\citet{Reynolds2010}}.
A feature near 7{\thinspace}keV was found, however.
Even though it was apparent from the
PN spectrum that the feature seems to be an edge, the
feature is very well fitted by a {\tt Gauss} model in {\it XSPEC}. The
centroid energy is 7.11{\thinspace}keV, which is exactly the energy of
the iron absorption K-edge reported by Suzaku. We proceed as in the 
Suzaku study, modeling this edge using the {\tt zvfeabs} model in
{\it XSPEC}. The fit, in this case, returns an iron absorption K-edge
of 7.19{\thinspace}keV. Freezing the edge energy to 7.11 returned a
$\chi^2_{\hbox{\tiny{red}}}$ value of 1.4, which is not as good as the
value obtained with the {\tt gauss} model (1.2). Notwithstanding this
difference in goodness of fit, it is our interpretation on physical
grounds that this is the iron absorption K-edge observed before by
Suzaku (\citeads{Reynolds2010}). To our knowledge, this is the
first detection by XMM of this feature in 1E{\thinspace}1740.7$-$2942.

For the Comptonization component, our first attempts were with the
simplest thermal {\tt compTT} (\citeads{1994ApJ...434..570T}) model in 
{\it XSPEC}, since this form of modeling was widely used in past spectra modeling
of 1E{\thinspace}1740.7$-$2942, and for others black hole
binaries and in a previous version of this study
(\citeads{Castro2012}). The spectra resulting of those fits are shown
in Figure{\thinspace}({\ref{fig1}}).

All the fits with this thermal Comptonization provided a very adequate
description of the spectrum. In our fits we kept the geometry
parameter equal to 1, which corresponds to a disk geometry in 
{\tt compTT}. Fits parameters can be found in 
Table{\thinspace}({\ref{tab2}}). 

Motivated by the evidence of Compton reflection found by some authors
in 1E{\thinspace}1740.7$-$2942 (\citeads{DelSanto05}), and by
the study of \citet{Natalucci2013}, which found no evidence of it, we
also used the convolutive reflect component in our fits
(\citeads{1995MNRAS.273..837M}). Presence of reflection is
very common in the low/hard spectra of black hole binaries (see, e.g.,
\citeads{1999MNRAS.303L..11Z}).  If the Comptonizing plasma is
surrounding the disk, then the presence of Compton reflection is
unavoidable. The results of our first attempt in modeling such
reflection, with the reflect model (acting on the
{\tt compTT} component) is shown in Table{\thinspace}({\ref{tab2}}). We found
indications of the possible presence of the Compton reflection in the 2003
and 2005 spectra but not in the 2012 spectrum. It is interesting to
note that the absence of reflection in 2012 and the presence of an
Fe-edge agrees with the results of \citet{Reynolds2010}.  As
can be see in Table{\thinspace}({\ref{tab2}}), the presence or
absence of the Compton reflection does not alter the parameters of
the fit with respect to the {\tt compTT} fit alone (i.e., without the
convolutive reflect component).

We have also made an attempt to fit our spectra with the {\tt compPS} model (\citeads{1996ApJ...470..249P}) in 
{\it XSPEC} to check what has been tried
before (\citeads{DelSanto05}). {\tt CompPS} 
provides a numerical solution of the radiative transfer equation, and 
it comprises Compton reflection as one of its parameters.
However, we noted that the 2012 spectrum in our fits did not fit very well by this model in the sense that it returns
an unrealistic value for the plasma temperature when using a slab
geometry with an optical depth around the value of one, as is the
case for our 2003 and 2005 fits with {\tt compTT}. Therefore, we decided not
to use the {\tt compPS} model in this study.

\begin{figure}[ht]
\resizebox{\hsize}{!}{\includegraphics[angle=0]{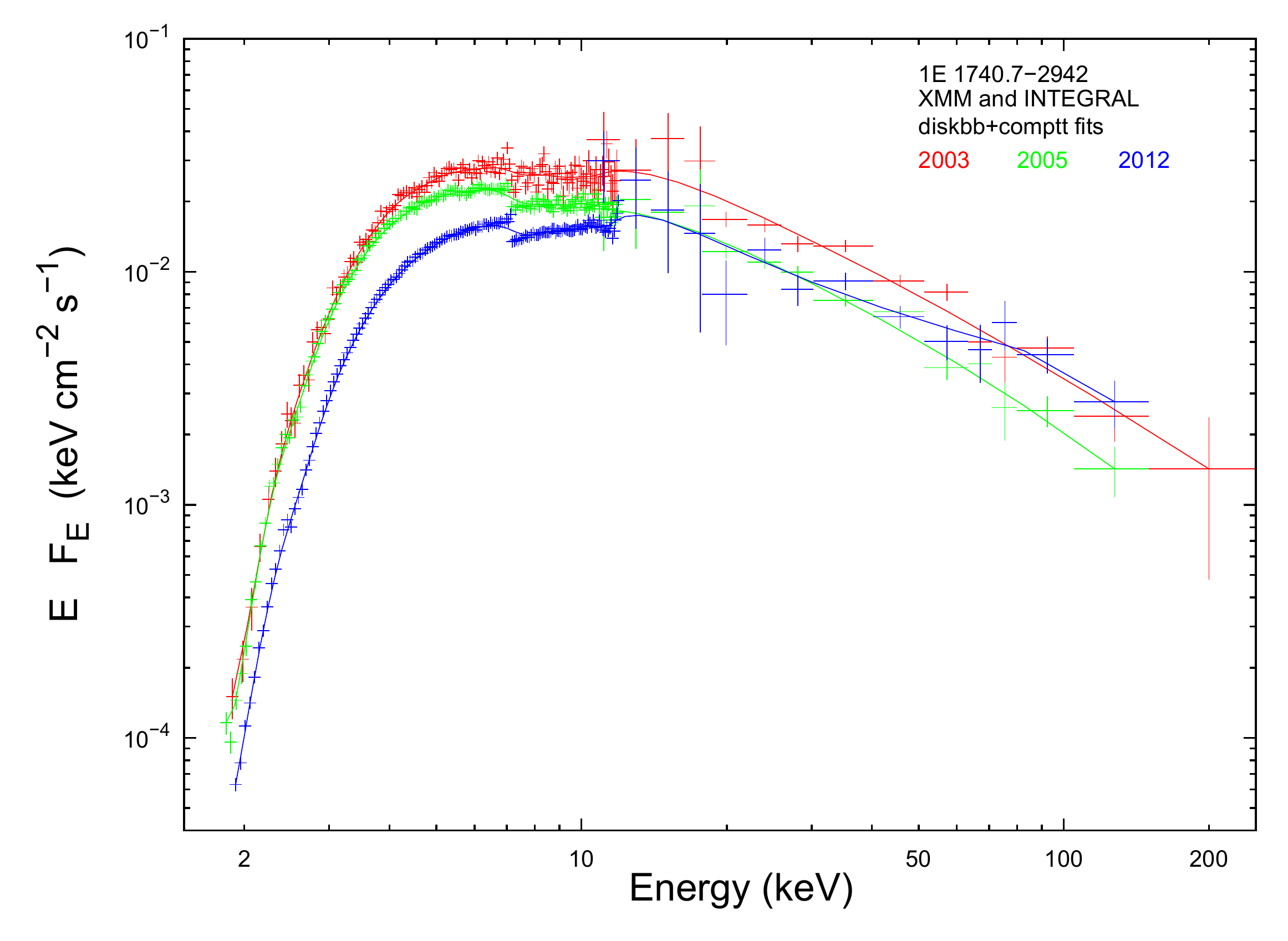}}
         \caption{
                  Absorbed spectra of 1E{\thinspace}1740.7$-$2942, which are modeled with
                  {\tt diskbb + compTT}, for the three XMM + INTEGRAL observations shown 
                  in this study. In only the 2012 spectrum, a feature at 
                  7.11{\thinspace}keV (modeled as a gauss in XSPEC), interpreted 
                  as being a Fe-edge (see text for details) is also included. 
                 } 
\label{fig2}
\end{figure}

In Figure{\thinspace}({\ref{fig2}}),
we show the spectral variation between the 2003,
2005, and 2012 spectrum.

We also show the parameters of our fits
to a exponential folded powerlaw model ({\tt cutoffpl} in {\it XSPEC}) in Table{\thinspace}({\ref{tab2}}).

In Table{\thinspace}({\ref{tab3}}), we show the measured fluxes for the
observations in this study.  It is interesting to compare our
broadband fluxes with those obtained by Suzaku
(\citeads{Reynolds2010}). The average flux measured by Suzaku in two
observations (one in $2006$ and another in $2008$) is $2.2 \times
10^{-9}${\thinspace}erg{\thinspace}cm$^{-2}${\thinspace}s$^{-1}$ in
the $2$--$300${\thinspace}keV range. For our observations, the average
between $2003$ and $2005$ observations is $2.5 \pm 0.4 \times
10^{-9}${\thinspace}erg{\thinspace}cm$^{-2}${\thinspace}s$^{-1}$ in
the same energy range ($2.97 \pm 0.1\times10^{-9}$ in 2003 and $1.97
\pm 0.2\times10^{-9}$ in 2005, in
erg{\thinspace}cm$^{-2}${\thinspace}s$^{-1}$), whereas our 2012 value
is $1.7 \pm 0.1 \times
10^{-9}${\thinspace}erg{\thinspace}cm$^{-2}${\thinspace}s$^{-1}$
(unabsorbed fluxes), characterizing a significant decrease in the
$2$--$300${\thinspace}keV flux from 1E{\thinspace}1740.7$-$2942 from
2003 up to 2012.

It is also interesting to compare our results for the hydrogen column
density ($N_{\mathrm{H}}$) with other studies. Observations with ASCA
\citep{1999ApJ...520..316S} found an average value of $9.7 \pm 0.4$ in
eight observations. A Chandra/HETGS study found $11.8 \pm 0.6$
(\citeads{2001ApJ...548..394C}), whereas a Chandra/ACIS-I found
$10.5 \pm 0.6$ (\citeads{Gallo2002}). The average in our three
observations is $13.6 \pm 0.1$. The average of all of these results is
$13.3 \pm 0.1$. Without our results, the average is $10.4 \pm 0.3$ with
all quoted values in units of $10^{22}${\thinspace}cm$^{-2}$. We
caution the reader that we performed our fits leaving $N_{\mathrm{H}}$ as a
free parameter. We then carefully verified that neither our
conclusions nor the quality of our fits is changed noticeably by adopting
(and freezing) $N_\mathrm{H}$ to $10.4$.

%for FD and FD alone: set font at emacs to size 7 to edit this table
%
\begin{table*}
\setlength{\extrarowheight}{1ex}
\caption{
         Model parameters of fitting XMM and INTEGRAL 1E{\thinspace}1740.7$-$2942 spectra.
        }
\label{tab2}
\begin{center}
\begin{tabular}{ccccc}
\hline \hline
\multicolumn{5}{c}{\bf{diskbb + compTT}}\\ \hline
 	                       &                                       &      2003 	        &       2005	               &        2012	            \\ \hline
{\tt phabs}                    & $N_{\mathrm{H}}$ ($10^{22}$ cm$^{-1}$) & 14.1$^{+0.3}_{-0.3}$   & 12.5$^{+0.1}_{-0.1}$        & 14.1$^{+0.1}_{-0.1}$        \\ \hline
{\tt diskbb}                   & $T_{in}$ (keV)                         & 0.24$^{+0.02}_{-0.02}$    & 0.17$^{+0.03}_{-0.03}$         & 0.19$^{+0.01}_{-0.01}$         \\ \hline
                               & $T_{0}$  (keV)                         & = $T_{in}$	        & = $T_{in}$                   & = $T_{in}$ 	              \\	
{\tt compTT}                   & $kT$ (keV)                            & 65.6$^{+2.2}_{-1.9}$   & 65.7$^{+1.5}_{-1.9}$        & 20.1$^{+0.1}_{-0.1}$         \\
  	                       &{\tt $\tau$}                           & 0.90$^{+0.08}_{-0.02}$    & 0.85$^{+0.02}_{-0.02}$         & 3.56$^{+0.03}_{-0.03}$          \\ \hline
{\tt gauss\tablefootmark{(a)}} &{\tt LineE} (keV)                      &	--              & --                          & 7.11$^{+0.01}_{-0.01}$          \\ \hline
${\chi}^2$/dof&                & 308/277                               & 400/294                & 352/298                                                    \\
\hline \hline
\multicolumn{5}{c}{\bf{diskbb + reflect*compTT}}\\ \hline
\tt{phabs}                     & $N_\mathrm{H}$ ($10^{22}$ cm$^{-2}$)     & 13.4$^{+0.4}_{-0.3}$   &	12.5$^{+0.1}_{-0.1}$  & 14.0$^{+0.1}_{-0.1}$         \\ \hline
\tt{diskbb}                    & $T_{in}$(keV)                          & 0.21$^{+0.03}_{-0.03}$    & 0.16$^{+0.01}_{-0.01}$         & 0.17$^{+0.01}_{-0.01}$	      \\ \hline	
\tt{reflect}                   &${\Omega}$/$2{\pi}$                    & 0.74$^{+0.29}_{-0.28}$	& 0.32$^{+0.15}_{-0.14}$         & $\leq$\,0.01\tablefootmark{(b)}         \\ \hline
\multirow{4}{*}{\tt{compTT}}   &$T_{0}$                                 & =$T_{in}$              & =$T_{in}$                    & =$T_{in}$        	      \\
 			       &$kT$(keV)                              & 64.7$^{+2.6}_{-2.4}$  & 65.5$^{+2.1}_{-2.0}$         & 20.8$^{+0.4}_{-0.4}$          \\
 			       &${\tau}$                               & 0.88$^{+0.05}_{-0.05}$   & 0.80$^{+0.03}_{-0.03}$          & 3.52$^{+0.03}_{-0.03}$           \\
 			       &$norm$      & 44$^{+3}_{-3}\times10^{-4}$ & 43$^{+7}_{-5}\times10^{-4}$ & 76$^{+2}_{-2}\times10^{-4}$	       \\ \hline
{\tt gauss\tablefootmark{(a)}} & {\tt LineE} (keV)                     & --                     & --                           & 7.11$^{+0.01}_{-0.01}$          \\ \hline
 $\chi^2$/dof		       &		                       & 296/275                & 394/292                      & 347/295	               \\
\hline \hline
%
% \multicolumn{5}{c}{\bf{diskbb + compPS}} \\ \hline
% \tt{phabs}		       & $N_\mathrm{H}$ ($10^{22}$ cm$^{-2}$)     & 13.43$^{+0.36}_{-0.32}$   & 12.42$^{+0.11}_{-0.11}$		&	13.83$^{+0.06}_{-0.06}$   \\ \hline
% \tt{diskbb}		       & $T_{in}$			       & 0.22$^{+0.03}_{-0.03}$	& 0.17$^{+0.01}_{-0.01}$		&	0.17$^{+0.01}_{-0.01}$	\\ \hline
%                                &$kT_{e}$ (keV)			       & 83.66$^{+3.52}_{-3.31}$	& 82.16$^{+4.76}_{-4.36}$		&	167.46$^{+3.05}_{-3.46}$	\\
%  			       &$kT_{bb}$ (keV)			       & =$T_{in}$		& =$T_{in}$			& =$T_{in}$			\\
% {\tt{compPS}}		       &${\mathcal{\tau}_{\scriptscriptstyle{T}}}$\tablefootmark{(b)} & 1.00 & 1.00				& 1.00				\\
%  			       &${\Omega}$/$2{\pi}$		       & 0.68$^{+0.28}_{-0.25}$	& 0.32$^{+0.02}_{-0.02}$		&	1.33$\times10^{-14}$	\\
%  			       &$norm$	  & 28.9$^{+15.8}_{-9.1}\times10^{3}$ & 62.7$^{+1.5}_{-1.5}\times10^{3}$ & 35.3$^{+6.9}_{-3.3}\times10^{3}$	                \\ \hline
% {\tt gauss\tablefootmark{(a)}} &{\tt LineE} (keV) 		       &	--		& --	                        & 7.11$^{+0.01}_{-0.01}$           \\ \hline
% $\chi^2$/dof		       &				       &	305/276		&	4712/293		&	390/298			\\
% \hline \hline
%
\multicolumn{5}{c}{\bf{diskbb + cutoffpl}}\\ \hline 
\tt{phabs}		       & $N_\mathrm{H}$ ($10^{22}$ cm$^{-2}$) & 13.0$^{+0.4}_{-0.3}$	& 12.3$^{+0.1}_{-0.1}$		& 13.8$^{+0.1}_{-0.1}$		\\ \hline
\tt{diskbb}		       & $T_{in}$(keV)			       & 0.21$^{+0.03}_{-0.03}$	& 0.15$^{+0.04}_{-0.04}$	& 0.17$^{+0.01}_{-0.01}$		\\ \hline	
                               & $\Gamma$			       & 1.42$^{+0.04}_{-0.04}$	& 1.52$^{+0.02}_{-0.02}$	& 1.24$^{+0.02}_{-0.02}$		\\
{\tt{cutoffpl}}		       & $E_{cut}$(keV)			       & 89.7$^{+15.7}_{-12.1}$	& 87.1$^{+14.5}_{-11.4}$	& 138$^{+51.7}_{-30.1}$	\\
 			       & $norm$	 & 85$^{+7}_{-6}\times10^{-3}$	& 81$^{+3}_{-3}\times10^{-3}$		& 333$^{+8}_{-8}\times10^{-4}$	\\ \hline
{\tt gauss\tablefootmark{(a)}} & {\tt LineE} (keV)                     &	--		& --	                        & 7.11$^{+0.01}_{-0.01}$           \\ \hline
 $\chi^2$/dof			&				       &	286/276		&	379/293			&	350/296			\\
 \hline \hline
 \end{tabular}
 \end{center}
 \tablefoot{
            Quoted errors are 1$\sigma$.
             \tablefoottext{a}{
                               This (additive) component is necessary to fit the 2012 spectrum (see text
                               for details).
                              }
 	     \tablefoottext{b}{3$\sigma$ upper limit in the parameter.}
           }
 \end{table*}
%
%endoftable2
%

%for FD and FD alone: set font at emacs to size 7 to edit this table
% 
\begin{table*}
\setlength{\extrarowheight}{2ex}
\caption{
         XMM and INTEGRAL measured (absorbed) fluxes for 1E{\thinspace}1740.7$-$2942 
         in erg{\thinspace}cm$^{-2}${\thinspace}s$^{-1}$ (errors are 1$\sigma$).
        }
\label{tab3}
\begin{center}
\begin{tabular}{ c c c c c c }
\hline \hline
        &		&  \multicolumn{1}{c}{2 -- 10 keV}    & \multicolumn{1}{c}{10 -- 20 keV}    & 20--50 keV                          & 50--200 keV             \\ \cline{2-6}
        & {\tt diskbb} 	& 1.15$^{+0.06}_{-0.06}$ $\times10^{-12}$ &           0            	    &    0 	                          &  0                                  \\
2003	& {\tt compTT}  & 2.74$^{+0.14}_{-0.14}$ $\times10^{-10}$ & 3.72$^{+0.19}_{-0.19}$ $\times10^{-10}$ & 5.65$^{+0.28}_{_0.28}$ $\times10^{-10}$ & 1.02$^{+0.05}_{-0.05}$ $\times10^{-9}$	\\
        & total        	& 2.76$^{+0.14}_{-0.14}$ $\times10^{-10}$ & 3.72$^{+0.19}_{-0.19}$ $\times10^{-10}$ & 5.65$^{+0.28}_{-0.28}$ $\times10^{-10}$ & 1.02$^{+0.05}_{-0.05}$ $\times10^{-9}$	\\ \hline

        & {\tt diskbb}  & 1.18$^{+0.20}_{-0.20}$ $\times10^{-13}$ &           0             	    &     0                		  &     0                     		 \\
2005    & {\tt compTT}  & 2.57$^{+0.28}_{-0.26}$ $\times10^{-10}$ & 2.56$^{+0.28}_{-0.26}$ $\times10^{-10}$ & 3.77$^{+0.41}_{-0.38}$ $\times10^{-10}$ & 6.44$^{+0.71}_{-0.64}$ $\times10^{-10}$  \\
        & total         & 2.57$^{+0.28}_{-0.26}$ $\times10^{-10}$ & 2.56$^{+0.28}_{-0.26}$ $\times10^{-10}$ & 3.77$^{+0.41}_{-0.38}$ $\times10^{-10}$ & 6.44$^{+0.71}_{-0.64}$ $\times10^{-10}$ \\ \hline
        & {\tt diskbb}  & 3.16$^{+0.02}_{-0.01}$ $\times10^{-13}$ &           0              	    &    0                     		  &      0                    		 \\
2012    & {\tt compTT}  & 1.86$^{+0.02}_{-0.01}$ $\times10^{-10}$ & 2.76$^{+0.01}_{-0.01}$ $\times10^{-10}$ & 3.89$^{+0.02}_{-0.01}$ $\times10^{-10}$ & 6.11$^{+0.01}_{-0.01}$ $\times10^{-10}$  \\
        & total         & 1.86$^{+0.02}_{-0.01}$ $\times10^{-10}$ & 2.76$^{+0.01}_{-0.01}$ $\times10^{-10}$ & 3.89$^{+0.02}_{-0.01}$ $\times10^{-10}$ & 6.11$^{+0.01}_{-0.01}$ $\times10^{-10}$  \\ 
 \hline \hline
 \end{tabular}
 \end{center}
 \end{table*}
 %
%endoftable3
%
%
%Figure2Begin
%author     : Flavio D'Amico
%short title: E F_E Spectra of 1E 1740.7-2942
%source     : made to appear here in this study, after referee's 1st comments
%Figure2End
%
%_________________________________________________________________
%
\section{Discussion}
\label{discussion}

The observations we describe in this study are one of the 
few  with broadband ($2$--$200${\thinspace}keV)
coverage of 1E{\thinspace}1740.7$-$2942 spectra with no
data gaps, which is provided only by imaging instruments,
avoiding any source confusion and, thus, flux contamination.
In our study, we tested the thermal--Compton paradigm for
the source spectra by  following, for example, other studies of 
1E{\thinspace}1740.7$-$2942 spectra. Such a paradigm was first
tested in Cyg{\thinspace}X$-$1 spectral analysis 
(\citeads{Gierlinski}). \\

For the first time to our knowledge, we reported here evidence
of a 7.11{\thinspace}keV Fe-edge detected by XMM, which  agrees
with previous reports for such a feature derived from Suzaku 
studies (\citeads{Reynolds2010}). \\

Our results for the magnitude of $N_{\mathrm{H}}$ for
1E{\thinspace}1740.7$-$2942 show a higher value than the
average. However, we do not claim here a variation of
$N_{\mathrm{H}}$ in the line of sight of 1E{\thinspace}1740.7$-$2942,
as well as any intrinsic change at or nearby the source, since our fits are
not very sensitive to the value of $N_{\mathrm{H}}$. \\

Our model of 1E{\thinspace}1740.7$-$2942 spectra shows a variation
in the optical depth of the plasma between 2012 when compared to 2003
or 2005.  This variation implies a change in the so-called $y$ parameter
of inverse Compton scattering (see, e.g.,
\citeads{1986rpa..book.....R}) on the order of 3.  From our database, it 
is not clear what causes this variation.  More simultaneous
observations by XMM and INTEGRAL will be very important in shedding
light in this issue. \\

Compared to the Suzaku study of 1E{\thinspace}1740.7$-$2942
(\citeads{Reynolds2010}), we verified a possible decrease in the
2--300{\thinspace}keV unabsorbed flux of the source. Our average value found for
2003 and 2005 is
$2.5{\pm}0.4{\times}10^{-9}${\thinspace}erg{\thinspace}cm$^{-2}${\thinspace}s$^{-1}$,
and our measured value for 2012 is
$1.7{\pm}0.1{\times}10^{-9}${\thinspace}erg{\thinspace}cm$^{-2}${\thinspace}s$^{-1}$,
while the averaged between 2006 and 2008 value measured by Suzaku is
$2.2{\times}10^{-9}${\thinspace}erg{\thinspace}cm$^{-2}${\thinspace}s$^{-1}$.
It is noteworthy that recent monitoring by the INTEGRAL/IBIS program
in the Galactic Bulge reported that 1E{\thinspace}1740.7$-$2942 is
below the detection sensitivity limit (\citeads{2013ATel.5332....1C}). \\

From the results in Table{\thinspace}({\ref{tab3}}), we can clearly see
a decrease in the $50$--$200${\thinspace}keV flux. The absorbed flux
in the $2$--$10${\thinspace}keV range remained within errors
constant in 2003 and 2005 and then decreased in 2012. In trying to
associate the change in the $50$--$200${\thinspace}keV flux with some
parameter of our models, no clear evidence is found, i.e., our
decrease in flux is not evidently correlated with any parameter. This
is contrary to what was observed in GX{\thinspace}339$-$4, where a
decrease in the luminosity is associated with an increase in
$kT_{\hbox{\tiny{e}}}$ (\citeads{2002MNRAS.337..829W}). The high
energy $50$--$200${\thinspace}keV flux must be associated with the
accretion disk (or, for example, with a corona surrounding it), and, similarly
to other Galactic black holes, it is the brightest component of the
X-ray spectrum when the source is in the LHS. \\

The comparison of our derived plasma temperatures with other
broadband hard X-ray studies of 1E{\thinspace}1740.7$-$2942 must be
considered with caution. For example, our measured
temperatures are different from the ones reported by \citet{DelSanto05}, but this is probably due to the fact that
these authors have used the physical assumptions of the {\tt compPS} model, which was not used
here. Similarly, a different modeling was used by
\citet{Bouchet2009} and \citet{Reynolds2010}. On the other hand, low temperatures of
${\sim}${\thinspace}$20${\thinspace}keV, as in our models of the
2012 spectrum, were already reported, for instance, by the NuSTAR
recent study (\citeads{Natalucci2013}), which has made use of the
same models we applied (i.e., {\tt diskbb} and {\tt compTT}). We
caution the reader that the NuSTAR reported value for the optical
depth is of the order of 1.4, while it varies in our observations
(see Table{\thinspace}{\ref{tab2}}). \\

It is also noteworthy that our fits with the {\tt cutoffpl}
model also provided acceptable fits, implying that non--thermal
process may also explain the broadband 1E{\thinspace}1740.7$-$2942
behavior. We note that a model based on a jet emission for explaining
the hard X-ray spectrum of 1E{\thinspace}1740.7$-$2942 was ruled out
(\citeads{BoschRamon}).  A model consisting of two
thermal Comptonization components was already used
(\citeads{Bouchet2009}) as an alternative to non--thermal
processes. \\

In our spectral analysis for 2003 and 2005, we found the presence of
Compton reflection, which agrees  with previous reports for 1E{\thinspace}1740.7$-$2942
spectra (e.g., \citeads{DelSanto05}).  Our 2012 spectral modeling,
however, found no evidence for such a component, which is also compatible with
other results (e.g., \citeads{Natalucci2013}).  It is very interesting
to note that in 2012 the observation of a Fe-edge in our spectrum
without the presence of Compton reflection is also agrees with
previous spectral analysis of 1E{\thinspace}1740.7$-$2942
(\citeads{Reynolds2010}).  Since the Compton reflection seems not
to be permanent in 1E{\thinspace}1740.7$-$2942 spectra, it is tempting to associate the vanishing of such a feature with
physical changes in possible corona surrounding the accretion
disk.  \\

Our study has revealed a rich spectral variability in 
1E{\thinspace}1740.7$-$2942 by highlighting the importance of broadband
coverage by XMM and INTEGRAL in future simultaneous observations. \\
 
%______________________________________________________________

\section{Conclusions}
\label{conclusion}

We have shown a simultaneous broadband study of
1E{\thinspace}1740.7$-$2942 in three different epochs with the use of
ESA's XMM and INTEGRAL satellites here that covers the band from $\sim${\thinspace}2 up to 200{\thinspace}keV with
no data gaps. The imaging instruments onboard XMM and INTEGRAL prevent
any kind of source-confusion/flux contamination. To our knowledge for the first time, we reported here a XMM/PN observation of the iron
absorption K-edge at $7.11${\thinspace}keV, a value reported previously in
Suzaku studies. We derived an historical decrease in the 
$2$--$300${\thinspace}keV flux of 1E{\thinspace}1740.7$-$2942. 

Our study has revealed a rich spectral variability in 1E{\thinspace}1740.7$-$2942. 
We have shown that the plasma temperature has varied between
the 2003--2005 and the 2012 spectra, but it is unclear from our analysis what 
the causes of this are. We note that this variation is accompanied
by a huge increase in the optical depth from 2003 to 2012.

We believe that only more broadband observations of
1E{\thinspace}1740.7$-$2942, for example, with simultaneous XMM and INTEGRAL
campaigns as the ones discussed in this study, will be able to
provide a data base from where tight constrains can be derived to the
Comptonization model for the emission in the LHS state of
1E{\thinspace}1740.7$-$2942.

%________________________________________________________________________

\begin{acknowledgements}
MC gratefully acknowledges CAPES/Brazil for support. MC
gratefully acknowledges Mariano M\'endez at the Kapteyn 
Astronomical Institute  (Groningen-The Netherlands) host 
in a visit in the period of September - October/2012,
sponsored by the COSPAR Program for Capacity Building 
Fellowship, for helpful discussions in the subject of 
this study. MC and FD acknowledges Raimundo Lopes de 
Oliveira Filho for early guidance in XMM data analysis 
issues. FD acknowledges Tomaso Belloni for helpful 
discussions. We deeply acknowledge Andrzej
Zdziarski, our referee, for superb comments and for 
helping us in improving the  quality of this study.
\end{acknowledgements}

%________________________________________________________________________
\bibliographystyle{aa}    %% bibliography style file aa.bst from A&A
\vspace{-9mm}
\bibliography{our_refs}   %% contains Bibtex entries copied from ADS
%________________________________________________________________________

\end{document}